\documentclass[onecolumn]{aastex631}

\newcommand{\eqq}[1]{Equation~(\ref{#1})}
\newcommand{\ie}{\textit{i.e.\/}}
\newcommand{\eg}{\textit{e.g.\/}}

\newcommand{\vecI}{\mathbf{I}}
\newcommand{\vecb}{\mathbf{b}}
\newcommand{\vece}{\mathbf{e}}
\newcommand{\bhat}{\mathbf{\hat b}}
\newcommand{\rhat}{\mathbf{\hat r}}
\newcommand{\phat}{\boldsymbol{\hat\phi}}
\newcommand{\uhat}{\boldsymbol{\hat u}}
\newcommand{\zhat}{\mathbf{\hat z}}
\newcommand{\vecp}{\mathbf{p}}
\newcommand{\vecq}{\mathbf{q}}
\newcommand{\vecr}{\mathbf{r}}
\newcommand{\vecv}{\mathbf{v}}

\newcommand{\matA}{A}
\newcommand{\matB}{B}

\newcommand{\nIr}{\left\langle nI_r^2\right\rangle}
\newcommand{\nIphi}{\left\langle nI_\phi^2\right\rangle}
\newcommand{\nIz}{\left\langle nI_z^2\right\rangle}

\newcommand{\Var}{\textrm{Var}}
\newcommand{\vcirc}{v_c}
\newcommand{\Msun}{M_\odot}
\newcommand{\covm}{C}
\newcommand{\lop}{\varpi}   
\newcommand{\ma}{\ell}  

\usepackage{natbib} 
\usepackage{amsmath}
\usepackage{enumitem}
\usepackage{verbatim}
\usepackage{graphicx}
\usepackage{subfigure}
\usepackage{color}
\usepackage{xcolor}
\usepackage{float}
\usepackage{hyperref}

\shorttitle{Asteroid Brownian motion}

\begin{document}

\title{Brownian motion of main-belt asteroids on human timescales} 

\author[0000-0002-8613-8259]{Gary M. Bernstein}
\affiliation{Department of Physics and Astronomy, University of Pennsylvania, Philadelphia, PA 19104, USA}
\email{garyb@physics.upenn.edu}
\correspondingauthor{Gary M. Bernstein}

\begin{abstract}
  The explosion of high-precision astrometric data on main-belt
  asteroids (MBAs) enables new
  inferences on gravitational and non-gravitational forces present in
  this region.  We estimate the size of MBA motions
  caused by mutual gravitational encounters with other MBAs that are
  either omitted from ephemeris models or have uncertain mass
  estimates. In other words, what is the typical Brownian motion among
  MBAs that cannot be predicted from the ephemeris, and therefore
  serves as noise on inferences from the MBAs?  We estimate the RMS
  azimuthal shift $\sigma_\phi$ of this ``Brownian noise''  by
  numerical estimation of the distribution of impulse sizes and
  directions among known MBAs, combined with analytical propagation
  into future positional uncertainties. At current
  levels of asteroid-mass knowledge, $\sigma_\phi$ rises to
  $\approx2$~km or $\approx1$~mas over $T=10$~yr, increasing as
  $T^{3/2},$  large enough to degrade many inferences from
  Gaia and LSST MBA data. LSST data will, however, improve MBA
  mass knowledge enough to lower this Brownian uncertainty by
  $\ge4\times.$  Radial and vertical Brownian noise  at $T=10$~yr are
  factors $\approx7$ and $\approx45,$ respectively,
  lower than the azimuthal noise, and grow as $T^{3/2}$
  and $T^{1/2}.$  For full exploitation of Gaia and
  LSST MBA data, ephemeris models should include the
  $\approx1000$~largest asteroids as active bodies with free masses,
  even if not all are well constrained.  This will correctly propagate
  the uncertainties from these 1000 sources' deflections into desired
  inferences. The RMS value of deflections from less-massive MBAs is
  then just $\sigma_\phi\approx60$~m or 30~$\mu$as, small enough to ignore until occultation-based position data become ubiquitously available for MBAs.
\end{abstract}


\section{Introduction}

High-precision measurements of positions of Solar System bodies have
great value to multiple scientific pursuits.  They can reveal
gravitational accelerations due to undiscovered bodies such as a
hypothetical ``Planet X'' \citep[\eg][]{holmanP9,inpopP9,trojans,occultations} or passing primordial
black holes \citep[\eg][]{pbh}, determine
the masses of known bodies \citep[\eg][]{goffin,baer}, test the laws
of physics \citep[\eg][]{inpopgraviton}, and characterize 
the non-gravitational forces that drive many forms of planetary
migration, such as the maintainence of the near-Earth asteroid (NEA)
population \citep[reviewed by][]{yarkovsky}.  While most applications of precision orbit determination
have used the 8 major planets as test particles, there are now $>10^6$
known minor planets available for this task.  These small bodies' astrometric
information content is growing extremely rapidly through large-scale
survey projects that can produce
milliarcsecond-scale accuracy for magnitude-limited small-body populations, such as Gaia \citep{gaiass3,gaiafpr},
PanSTARRS,\footnote{\url{https://www2.ifa.hawaii.edu/research/Pan-STARRS.shtml}}
and particularly the upcoming Rubin observatory
\textit{Legacy Survey of Space and Time} (LSST).\footnote{\url{https://rubinobservatory.org/}}  The great majority
of currently-tracked objects are main-belt asteroids (MBAs), so we
wish to investigate the potential power of MBAs as gravitational test
bodies.

The most advanced ephemerides for the Solar System \citep{de440,inpop,pitjeva} consider the
gravitational forces emanating from the Sun, from the major planets and their large
moons, from up to $\approx300$ of the largest individual MBAs (typically those that can influence the exquisitely measured ranging to Mars), and
from distributed mass annuli representing the smaller members of
 the asteroid belt(s) and Kuiper belt.
Here we ask: \textit{what is the RMS ephemeris error in angular or radial
position a typical MBA accrues because of its gravitational
encounters with other individual MBAs?}  We are interested not in the
perturbations that can be well described by a multipolar ring model
for the full population; rather the Brownian motion that accumulates
from closer encounters with: (a) MBAs that \emph{are not included} in the
ephemeris model at all; and (b) asteroids that \emph{are} included
in an ephemeris model, but have some uncertainty $\sigma_M$ in their masses.
Any such departures became a source of noise in any
inferences that we make from use of these MBAs as
dynamical tracer particles.  We aim to quantify this noise, as a
function of the RMS uncertainty $\sigma_M$ on the masses of individual
asteroids, and assuming that deflectors with $M<\sigma_M$ are not
included in the ephemeris at all.

In principal, one can collect all of the positional information for
all $>10^6$ known solar-system bodies, and fit them to a model with
a free state vector and mass for each body.  Calculating the influence
of $>7\times10^6$ free parameters on $10^8$ or so observables is,
however, unlikely to be computationally feasible, and most of the
bodies' masses will be lost in measurement noise and/or be degenerate
with other bodies' masses.  The most ambitious attempt at a global
solution that we are aware of is that of \citet{goffin}, who was able
to include 250 MBAs as gravitating bodies in the fit to the full
corpus of observations at the time.  For only $\approx 1/2$ of these
bodies did the fit yield masses $>3\times$ the uncertainties.  With
vast improvements in data and computing power since then and in the
next decade, the number of detectable
asteroid masses should increase, but there will always be
uncertainties that accrue from the unmodelled deflectors and the
finite $\sigma_M$ of modelled deflectors.  These remnant uncertainties
will in turn serve as noise in attempts to model new sources of
acceleration.

While long-term diffusion of orbital elements has been extensively
investigated in the context of planetary-system formation and
planetary migration, we address here the more limited question of how
much the MBAs will have their orbit elements and positions altered
by Brownian motion over a time period $T \lesssim100$~yr of human
observation, in the present dynamical environment.  Indeed essentially
all of the relevant deflector bodies and circumstances of encounters
are known.  We will make use
of the currently known population of MBAs, correcting for
incompleteness of current surveys at lower masses---or more precisely,
at fainter absolute magnitudes $H,$ since the $H$ distribution is
known far more accurately than the $M$ distribution.

\section{Calculation overview}
We aim to calculate the Brownian motion variance to an accuracy of a
factor $\approx 2.$
Greater accuracy in the calculation is not warranted
since some of the principal inputs have substantial uncertainty.  The
number of MBAs vs mass is 
uncertain because of unknowns in the conversion
from $H$ to mass; the status of future observations and $\sigma_M$ is
also unclear.  We are hence justified in retaining only the
contributions to our results that are leading
order in the orbital eccentricity $e$ of the tracer MBA.
We will assume throughout that the joint
distribution of mass (or $H$) and orbital elements is separable into a
mass distribution and an orbital distribution.  In this scenario, the
Brownian motion is independent of the tracer mass, and involves an
integral over the mass distribution of the deflecting bodies.

We adopt a cylindrical coordinate system $(r,\phi,z),$ with $\zhat$
normal to the initial orbital plane, and $\phi=z=0$ toward the
perhelion of the initial orbit, \ie\ both the initial ascending node
$\Omega_0$ and longitude of perihelion $\lop_0$ equal to zero.  The
unit vectors $\rhat, \phat$ rotate with the target asteroid.  The mean
anomaly at time $t=0$ is $\ma_0.$  In general, subscripts of $0$ will
indicate properties of the unperturbed orbit.

All distances will be given in units of the original semimajor axis $a_0$, and all velocities in units of the circular velocity $\vcirc \equiv \sqrt{G\Msun/a_0}.$  In these units, $G\Msun=1,$ the period of the initial orbit is $2\pi,$ and the (unperturbed) mean anomaly is $\ma=t+\ma_0.$ 

We describe all encounters with other asteroids in the impulse approximation, defining $\vecI$ as the $\Delta\vecv$ imparted on the target by the deflector.
In our units, the gravitational impulse imparted by a deflector of mass $M_d$ approaching at impact parameter $\vecb=b\,\bhat$ and relative velocity $v$ is
\begin{equation}
  \vecI = 2 \frac{M_d}{\Msun} (bv)^{-1} \bhat.
  \label{eq:impulse}
\end{equation}
All of our results will be derived at leading order in $\vecI,$ which
is very well justified by the small size of MBA-induced impulses.

One of our tasks will be to derive, from the known asteroid population, the rate (per tracer) of encounters vs the imparted impulse,
\begin{equation}
  \frac{dN}{dt\,dI_r\,dI_\phi\,dI_z},
  \label{eq:dN}
\end{equation}
where the components of $\vecI$ are given in the cylindrical basis
vectors about the asteroid's position at the impulse.
We will estimate this function from the known population,
approximating it as constant within each of three subsets of tracer MBAs:
the Inner, Middle, and Outer regions bounded by the 4:1, 3:1,
5:2, and 2:1 mean-motion resonances with Jupiter.
We will further assume that the impulses on a given target are drawn
independently from this distribution, \ie\ a Poisson process defined
by this rate.  In this case, the only properties of the impulse
distribution that we need are its second moments for components $\alpha \in \{r,\phi,z\}$
\begin{equation}
  \left \langle n I_\alpha^2 \right\rangle \equiv \int d^3I \frac{dN}{dt\,dI_r\,dI_\phi\,dI_z} I_\alpha^2.
\label{eq:nvsq}
\end{equation}
The average of the cross terms $I_rI_z$ and $I_\phi I_z$ vanishes if
the deflector distribution is symmetric in inclination as expected, and we find numerically that the mean $I_rI_\phi$ is small enough to ignore.

From this knowledge of the impulse distribution, our goal is to obtain
the covariance matrix of the deviations of the target's position from
the initial orbit, after some time $T.$ The position observables are
the range, plus the celestial latitude and longitude of the target.  We will simplify our results by assuming a heliocentric observer, so that the observational position vector is $\vecp\equiv (r,\phi,\theta=z/r).$ The quantity we seek is the covariance matrix $\covm^p$ of the observations attributable to the accumulated gravitational perturbations:
\begin{equation}
  \covm^p  \equiv  \left\langle  \Delta\vecp \Delta\vecp^T\right\rangle,  \label{eq:Cp}
\end{equation}
where the angle brackets indicate an average over possible
realizations of the impulse history, and over the mean anomaly $\ma$
at the time of observation.  To do so, we will introduce an
intermediate set of 6 orbital elements $\vecq,$ selected to respond linearly to $\vecI$ at $|I|\ll 1.$
We will derive the matrix $\matA$ that describes the orbital-element shifts at time $T$ that arise from an impulse at time $t_i$: 
\begin{equation}
  \Delta \vecq(e_0,T,t_i) = \matA(e_0, T, t_i) \, \vecI.
  \label{eq:A}
\end{equation}
In our first-order perturbation theory, $\Delta\vecq(T)$ will be the sum of the $\Delta\vecq_i$ imparted by all impulses applied at times $0<t_i<T.$  Because the impulses are uncorrelated, $\vecq$ will therefore be the result of a random walk.  The distribution will have a covariance matrix $\covm^q(T) \equiv \left\langle \Delta\vecq(T) \Delta\vecq^T(T) \right\rangle$ whose elements are
\begin{eqnarray}
  \covm^q_{jk}(e_o,T) & = & \left\langle \sum_{i,\gamma} \matA_{j\gamma}(e_0,T,t_i) \matA_{k\gamma}(e_0,T,t_i) I^2_{i,\gamma} \right\rangle \\
           & = & \sum_\gamma \int dt_i \matA_{j\gamma}(e_0,T,t_i) \matA_{k\gamma}(e_0,T,t_i) \left\langle n I_\gamma^2\right\rangle.
\label{eq:Cqjk}
\end{eqnarray}
In the first line, the sum $i$ runs over the impulses and $\gamma$ runs over the components $r,\phi,z$ of the impulse. The second line evaluates the expectation value of averaging over realizations of the random walk of impulses, exploiting the independence of the individual impulses from each other.

The last element of our calculation will be a conversion from the
element shifts $\Delta\vecq$ into the observed displacements $\Delta\vecp$
at time $T.$  In linear perturbation theory this will again be expressible as a matrix
\begin{eqnarray}
\label{eq:B}
  \Delta\vecp(e_0,T) & = & \matB(e_0,T) \vecq(T) \\
  \Rightarrow \quad 
  \covm^p_{\alpha\beta}(e_0,T) & = & \sum_{jk} B_{\alpha j}(e_0,T)
                                     B_{\beta k}(e_0,T) \, \covm^q_{jk}(T) 
\label{eq:covp}
\end{eqnarray}
which we will average over the phase of the MBA orbit at the
observation time $T,$ \ie\ average over mean anomaly $\ma.$

Section~\ref{sec:impulse} describes the estimation of $\left\langle
  nI_\alpha^2\right\rangle$ for MBA regions.
Section~\ref{sec:propagation} derives the forms of $\matA(e_0,T,t_i),$
and $\matB(e_0,T).$  Section~\ref{sec:results} combines these and
summarizes the results.  The implications of these results for use of
MBA observations for high-precision inferences are discussed in Section~\ref{sec:discuss}.

\section{Impulse distribution}
\label{sec:impulse}
\subsection{Derivation}
Under the assumption that the distributions of masses and orbital
elements of MBAs are separable, the quantities needed to describe the
random walk of the asteroid's orbits can be expressed as an integral
over the absolute magnitude $H$ of the deflectors, the relative velocity $v$,
impact parameter $b,$ and polar/azimuthal angles $\theta,\phi$ of the
unit vector of the impulse direction $\bhat$:
\begin{eqnarray}
  \left\langle n I_\gamma^2 \right\rangle & = & \int
                                              dH\,dv\,db\,d\theta\,d\phi
                                             \frac{dN}{dH} \frac{dn}{dv\,db\,d\theta\,d\phi}
                                              I^2_\gamma(H,v,b,\theta,\phi)
  \\
  & = & \int dH \frac{dN}{dH} \int dv
        \frac{dn}{dA\,dv\,d\theta\,d\phi} \hat b_\gamma^2(\theta,\phi)
        \int_{b_{\rm min}}^{b_{\rm max}} 2\pi b\,db\,
        \left(\frac{2\sigma_M(H)}{bv}\right)^2 \\
  & = & 8\pi \int dH \frac{dN}{dH} \sigma^2_M(H)  \int dv
        \frac{dn}{dA\,dv} v^{-2} \left\langle\hat b_\gamma^2(v)\right\rangle
        \log(b_{\rm max}/b_{\rm min}).
\end{eqnarray}
The second row adopts the (numerically verified) assumption that the
distribution of impact parameter $b$ will always be $\propto dA=2\pi
b\,db$ in the range of $b$ of interest to us.  The mean rate of encounters
between a deflector-tracer pair, as a function of the impact velocity
and geometry, is $dn/dA\,dv\,d\theta\,d\phi,$ which we will determine
through numerical measurements with the orbits of all known MBAs. The
$H$ distribution of MBAs is $dN/dH.$ We define $\sigma_M(H)$ via
\begin{equation}
  \sigma_M(H) \equiv \left\{\begin{array}{cl}
                              \sigma_M, & M(H) > \sigma_M \\
                              M(H),        & M(H) \le \sigma_M
                            \end{array} \right.
  \label{eq:sigmaMH}
\end{equation}
such that it represents the error in the ephemeris due to uncertainty
in the masses of MBAs included in the ephemeris (first row), and due
to the entirety of the masses of MBAs omitted from the ephemeris
(second row).

In the third line, we execute the integral over $b$ and introduce the
average geometry factors $\langle \hat b^2_\gamma\rangle$ of encounters as
a function of $v$.  It must be true that
\begin{equation}
  \left\langle \hat b_r^2 \right\rangle
  + \left\langle \hat b_\phi^2 \right\rangle
  + \left\langle \hat b_z^2 \right\rangle = 1,
\end{equation}
but equality of the three is not required.

For the values of $b_{\rm min}$ and $b_{\rm max},$ we adopt heuristics
for the nature of impulses of interest.  We set $b_{\rm max}$ by 
requiring that $b/v<1,$ such that the ``impulse'' is being applied
over a time shorter than $1/2\pi$ of the orbital period.  Events
lasting longer than this are either extended close encounters of 2
MBAs with very similar orbits---which are rare and can be identified
and modeled in advance; or they are slow, longer-range interactions
that would be adequately modeled by a multipole model of the
collective mass of the asteroid belts.  Thus we take $b_{\rm max}=v.$

For $b_{\rm min},$ we adopt the criterion that encounters at
sufficiently small $b$ will generate sufficiently large impulses $I$
that observations of the tracer will detect this individual impulse at
levels well above measurement noise.  This would mean that an
ephemeris model could include the mass of the deflector asteroid, and
this mass could be usefully determined from fitting the data of just this
single tracer.
Such encounters would thus no longer be considered source of
stochastic Brownian motion.  Finding all such encounters among known
MBAs is feasible, and in fact is forecasted for the next decade's LSST
observations by
\citet{negin}. 

If we crudely choose some threshold
$I_{\rm det}$ of impulse as being large enough to generate high-$S/N$
deflections on its tracer, then our condition becomes $2M(H)/bv <
I_{\rm det},$ or $b_{\rm min}=2M(H)/vI_{\rm det}.$  We then have
\begin{equation}
   \left\langle n I_\gamma^2 \right\rangle = 
8\pi \int dH \frac{dN}{dH} \sigma^2_M(H)  \int_{\sqrt{2M(H)/I_{\rm
      det}}}^\infty \frac{dv}{v^2} 
        \frac{dn}{dA\,dv}  \left\langle\hat b_\gamma^2(v)\right\rangle
        \log \left[v^2 I_{\rm det}/2M(H)\right].
        \label{eq:nI2}
      \end{equation}
The lower bound on the $v$ integral marks the speed below which
$b_{\rm max}<b_{\rm min}.$  Since $b_{\rm min}$ and $b_{\rm max}$
appear logarithmically, the estimated diffusion is relatively insensitive to choices of
their values.

\subsection{Numerical results}
\eqq{eq:nI2} reduces the calculation to a double integral over the
deflector $H$ and the tracer-deflector relative velocity $v.$  The
constituents of this calculation are estimated as follows.

\subsubsection{Encounter rates and geometries}
The pairwise interaction rate $dn/dA\,dv$ and the impulse direction
components $\langle \hat b^2_\gamma \rangle$ are estimated from a
numerical integration of all of the MBA orbits available
from the Minor Planet Center (MPC) as of 6 January 2025.   We retain as
potential deflectors the 1.26~million bodies with $1.8<a<4.2$~AU,
observations at multiple oppositions, and uncertainty values $U\le5.$
We designate each source as being an ``Inner'' MBA, with semi-major axes between the 4:1 and 3:1
mean-motion resonances of Jupiter; ``Middle'' MBA between the 3:1 and
5:2 resonances; ``Outer'' MBAs between the 5:2 and 2:1
resonances; the remaining ``Other'' objects are a small minority that
contribute very little to the total diffusion rate, and we will ignore.  Our working
assumption will be that the MPC objects are an unbiased sample of the
orbital-element distribution within each of the Inner, Middle, and
Outer belt regions.

From this full MBA list, we select a random subset of 5000 objects
from each of the Inner, Middle, and Outer belts to serve as a sample
of ``tracers.''  We record all of the passages of any tracer MBA
within 0.03~AU of any other ``deflector'' MBA (drawn from the full
catalog of 1.26~million) over a 10-year period.  The 5000 tracers per
belt region are enough to be representative of the statistics of
the region, but few enough that the calculation can be done easily
on a laptop computer.
We will estimate the constituents of  $\langle nI_\gamma^2\rangle$ in
\eqq{eq:nI2} separately for each of the nine combinations of (tracer region, deflector region).

Using a simple leapfrog integrator with gravity from the Sun and the 8
major planets, we advance all MBA's orbits from their heliocentric
osculating elements at the epochs given by the MPC, to barycentric
state vectors on 1 May 2025.  Using $kd$-trees to accelerate
pair-finding, we locate all tracer-deflector pairs
that pass within 0.03~AU of each other during the $\pm1$-day period
around this initial epoch, assuming inertial relative motion during
this interval.  The circumstances of such encounters are saved: the
identities of the two MBAs involved, the time of closest approach, the
relative velocity $v,$ and the impact parameter vector $\vecb.$

The leapfrog integrator advances all 1.26 million state vectors by 2
days, and the process is repeated.  We continue searching for pairs at
2-day intervals until we have recorded 10 years' worth of encounters---17.7 million
total impulses, or roughly 1000 encounters per tracer.

\begin{figure}
  \centering
  \includegraphics[width=0.8\textwidth]{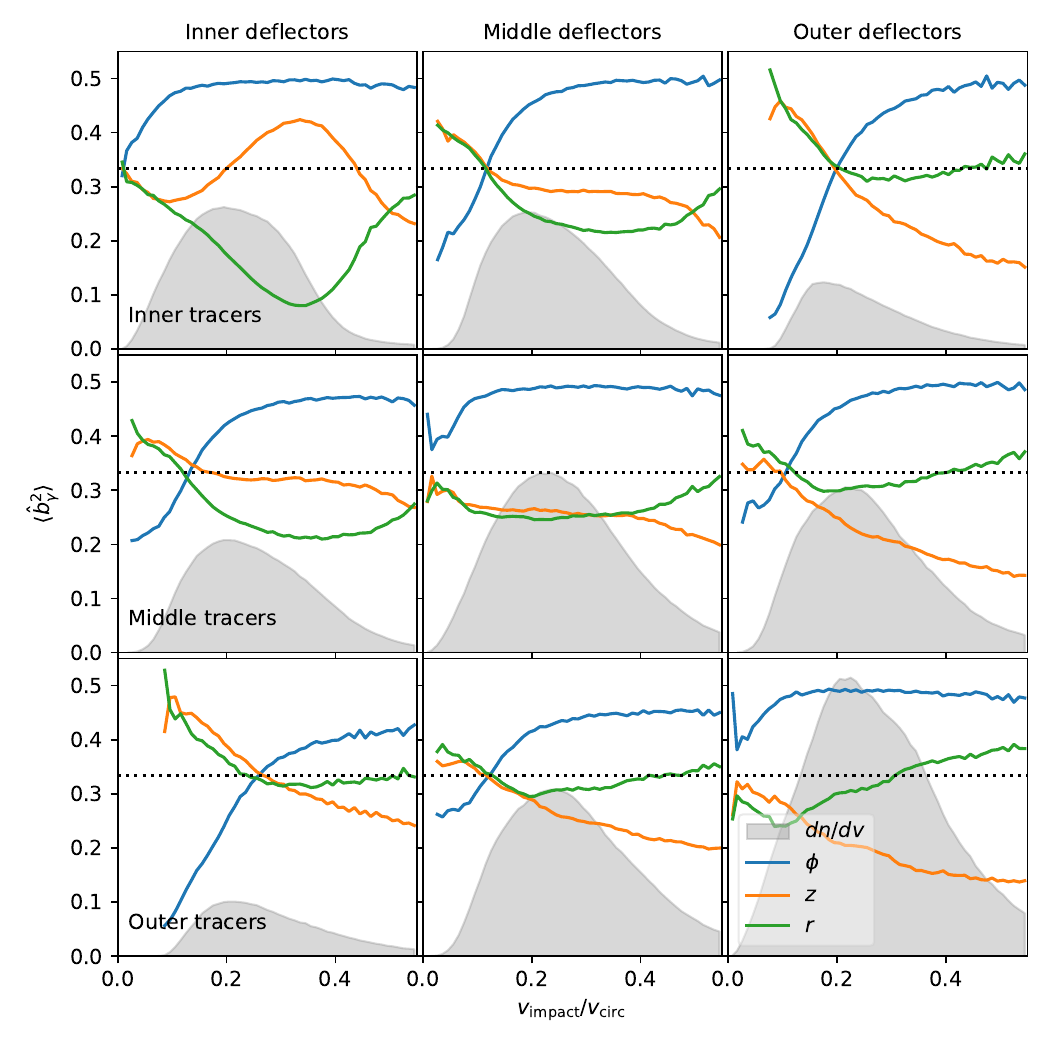}
  \caption{Each panel shows statistics of the encounters between
    tracer MBAs from one region and deflector MBAs from another.  As a
    function of the relative velocity at impact, the three curves show
    the mean squared value of the radial, azimuthal, and vertical
    components of the impulse direction $\bhat.$  The horizontal
    dashed line at value $1/3$ would be the result for isotropic
    encounters, but we see that the azimuthal component of the impulse
    tends to be larger, with a variety of behaviors.  The gray regions
    trace the distribution $dN/dv$ of the rate of interactions per
    tracer per orbit per interval in $v.$  These curves have a common
    (arbitrary) normalization, so the relative heights properly
    reflect the relative frequency of encounters among
    pairs of deflector and tracer regions.}
  \label{fig:rtz}
\end{figure}

For each pair of tracer-deflector regions (\eg\ Inner-Inner), we combine the list of
events with the counts of candidate tracers and deflectors to
calculate $dn/dA\,dv,$ the event rates per eligible pair; and the mean
geometry of the encounter, $\langle \hat b^2_\gamma \rangle,$ as a
function of $v$, all plotted in Figure~\ref{fig:rtz}.
Here it is apparent that the $\hat b_\phi$ component is, on average,
larger than $\hat b_z$ and $\hat b_r,$ but there is significant
variation with MBA class and with $v$.  An important result is that
the quantity $v^{-2} (dN/dA\,dv)$ is always well-behaved as $v\rightarrow 0,$
so we do not need any special treatment to sum the integrals in
\eqq{eq:nI2}.  The $v$ integrand peak is in the range $0.03<v/v_c<0.1.$

\subsubsection{Asteroid properties}
\label{sec:mbaprops}
We crudely approximate the population as having a geometric albedo of 0.25,
density of 2500~kg~m$^{-3},$ and spherical shapes, which leads to the
conversion
\begin{equation}
  M(H) = 1.2\times10^{-17} M_\odot \times 10^{-0.6(H-15)}
\end{equation}

\begin{figure}
  \centering
  \includegraphics[width=0.8\textwidth]{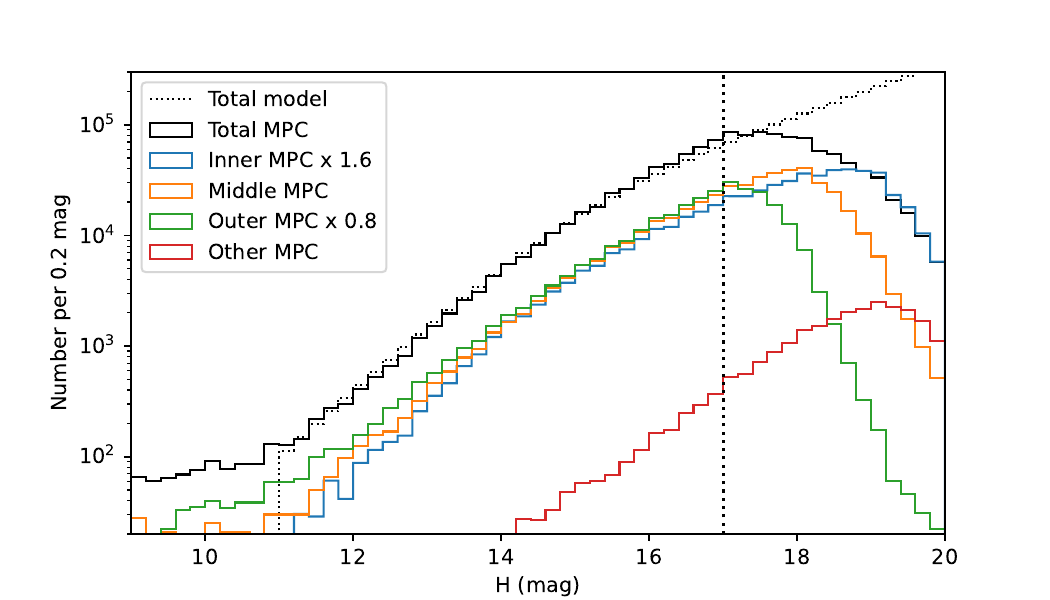}
  \caption{Differential counts of MBAs cataloged by the MPC are
    plotted vs absolute magnitude $H$  
    for the Inner, Middle, Outer, and ``Other'' regions of
    the main belt, and for their total.  We assume that they are
    complete for $H\le17,$ and extrapolate to fainter sources using
    the functional form from \eqq{eq:lsstmba} shown as the dotted
    curve. The Inner and Outer regions are slightly rescaled to help
    visual comparison of the curves' shapes in the three MBA regions.}
  \label{fig:counts}
\end{figure}
We also require the $H$ distribution of potential deflectors.
Figure~\ref{fig:counts} plots our assumptions for $dN/dH.$  The MPC
catalog is likely close to complete in all three regions for $H\le17,$
so we will use the MPC counts directly in this regime.  Note that the
three regions have similar shapes of $dN/dH.$  For $H>17,$ we adopt
the functional form for the cumulative MBA counts given by
initial LSST predictions \citep{lsstbook}:
\begin{equation}
  N(<H) \propto \frac{10^{0.43(H-15.7)}}{10^{0.18(H-15.7)} +
    10^{-0.18(H-15.7)}},
\label{eq:lsstmba}
\end{equation}
normalizing this curve to the observed $H<17$ counts for each region.

Our integral for the uncertainty due to unmodelled MBA gravitation
requires the RMS error $\sigma_M(H)$ between the true mass $M$ and the
mass used (if any) in the ephemeris model.
The two largest asteroids, Ceres and Vesta, have masses determined to
high precision by the \textit{Dawn} spacecraft \citep{Ceres,Vesta}.  Thus despite holding
half of the $\approx 10^{-9} M_\odot$ mass of the asteroid belt, their $\sigma_M$
values\footnote{More precisely, the uncertainties in their $GM$ values
  relative to $GM_\odot$.} are 
$\le3\times10^{-15} M_\odot$ and
this unknown part of their contribution to other MBA's motion is 
unimportant.  Aside from a handful of other asteroids in
well-characterized binary systems or visited by spacecraft, most other knowledge of MBA masses
has and will come from mutual gravitational encounters.   An
individual asteroid's mass can be estimated from its measured effect  on the
few MBAs on which it imparts the largest impulses.  The accuracy
$\sigma_M$ of this determination will vary, depending upon the
geometry  and timing of
each deflector MBA's encounters, and the measurement errors on its tracers.  These
factors are independent of the mass of the deflector.  We will
assign a value $\sigma_M$ to the RMS uncertainty in MBA mass
attainable from following a few individual tracers, and calculate
the Brownian-motion noise as a function of $\sigma_M.$

\begin{figure}
  \plottwo{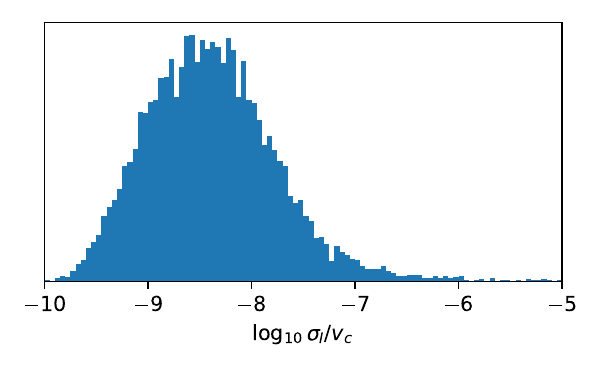}{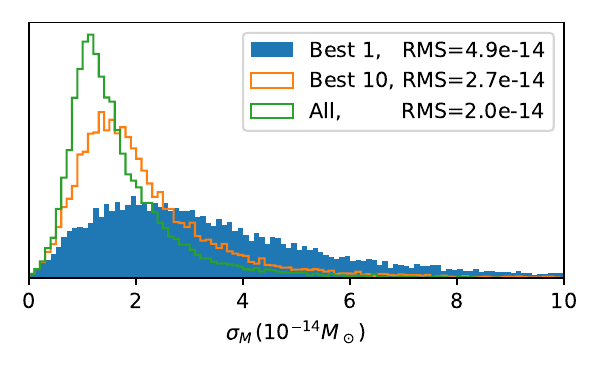}
  \caption{Forecasts of uncertainties derived from LSST tracking of the known MBAs.  The left-hand plot shows the distribution of (log) uncertainties $\sigma_I$ on the amplitude of the impulse applied to the tracer MBA, incorporating forecasted LSST observations and marginalizing over the tracer's initial state vector.  The right-hand side shows the distribution of the uncertainties on the mass, $\sigma_M=(bv/2)\sigma_I,$ for deflector MBAs expected from LSST.  The histograms show the results from using only the single tracer yielding the lowest $\sigma_M$ for that deflector; combining the 10 best-$\sigma_M$ tracers; or combining all impulses at $b<0.03$~AU.  The RMS value of $\sigma_M$ over the deflector population is given in the legend.  Masses are in solar units, and velocities in units of the circular velocity of the deflector.}
  \label{fig:sigmaI}
\end{figure}

To give some context, a
comprehensive investigation of MBA masses from mutual scattering
events by \citet{goffin} reports typical $\sigma_M\approx
5\times10^{-13}$ for 30 MBAs (although the list is biased toward
better-measured events). But we can expect substantial improvement
in the LSST era when a millions of potential tracer asteroids attain
mas-level astrometry. \citet{negin} use a simulated LSST sequence of
observations to estimate the impulse (and mass) uncertainties from
analyzing LSST's data on mutual encounters of the known asteroids. The
tracer MBAs for those events are essentially an unbiased selection
from the known MBAs, and the left-hand plot of Figure~\ref{fig:sigmaI}
shows the distribution of the uncertainty $\sigma_I$ derived from each
tracer's LSST data after marginalizing over its initial state vector.
This histogram marginalizes over the apparent magnitude and observing
conditions of the tracer, the date of the encounter, and the direction
$\bhat$ of the impulse.  The right-hand plot of
Figure~\ref{fig:sigmaI} forecasts the distribution of $\sigma_M$ that
we can expect for deflectors.  This forecast is made as follows: we
treat each of the 5000 tracer MBAs that we have selected from each
region as deflectors, since we have a list of all of their encounters
with known MBAs with $b<0.03$~AU.  For each such encounter, we select
at random one $\sigma_I$ from the distribution for LSST data on known
MBAs forecasted by \citet{negin} (Figure~\ref{fig:sigmaI}, left), and
calculate $\sigma_M=(bv/2)\sigma_I.$  For each deflector, we then
calculate (a) the individual encounter with lowest $\sigma_M$; (b) the
combined constraint $\left(\sum \sigma_{M,i}^2\right)^{-1/2}$ from the
10 lowest $\sigma_M$ encounters for that deflector; and (c) the
combined constraint for all $b<0.03$~AU encounters (typically
$\sim1000$ of them).  The right-hand plot of Figure~\ref{fig:sigmaI}
gives the histograms of these three measures of $\sigma_M.$  We see
that most of the available information on any given deflector's mass
comes from its 10 most-informative encounters.  It should not be a
computational barrier to solve for the masses of $O(10^4)$ MBAs by including the deflections they impart on $O(10^5)$ other MBAs in an ephemeris fit.
The RMS $\sigma_M$ using LSST data from the 10 most-informative tracer encounters with a given deflector is estimated as $\sigma_M=2.6\times10^{-14}M_\odot.$ There are $\approx450$ MBAs with mass above this forecasted RMS $\sigma_M.$  This is a conservative estimate in that we ignore any observational information before LSST, and we also do not consider the information increase that will come from tracking the $\approx10$-fold increase in number of known MBAs that LSST will enable.

\subsubsection{Individually detectable impulses}
Examining the $\sigma_I$ distribution forecasted for LSST data on
tracers by \citet{negin}, we find that for impulses occurring near the midpoint of the survey, $\approx80\%$ of all cases yield $\sigma_I<10^{-4}\,{\rm m}\,{\rm
  s}^{-1},$ some an order of magnitude lower.  We conservatively
assume that any encounter producing  an impulse $>5\times$ this value, $I_{\rm det}=5\times 10^{-4}\,{\rm m}\,{\rm
  s}^{-1},$ can be used to directly constrain the relevant deflector and include it in the ephemeris model.
When normalized by $v_c,$ we have $I_{\rm  det}=10^{-8}$ at 10\% accuracy across the main belt.

\section{Propagation to position shifts}
\label{sec:propagation}

\subsection{Element shifts from impulses}
\label{sec:elements}
The orbital shifts from an impulse can be derived from Gauss's
equations, \eg\ as presented by \citet[][Section 1.9.2]{tremaine}, but
we provide here a derivation customized to our case.
Five of our orbital elements $\vecq$ will be taken from constants of the motion.  The mean anomaly $\ma$ is not appropriate as the sixth, time-dependent element of $\vecq$ because the derivative $d\ma/dI$ can diverge as $e_0\rightarrow 0.$ Instead we introduce $\tau = \ma+\lop,$ which we will show does have a shift $\Delta\tau$ during an impulse that has finite and linear response to $\vecI$.  Between impulses, $\tau$ advances with the mean motion as $a^{-3/2}t.$  We define a coordinate system that is uniformly rotating with the unperturbed $\tau_0=t+\ma_0.$ The two smoothly rotating unit vectors $\uhat_\parallel=(\cos \tau_0, \sin \tau_0)$ and
$\uhat_\perp=(-\sin \tau_0, \cos \tau_0)$ satisfy $\uhat_\parallel
\times \uhat_\perp = \zhat,$ and we will use subscripts $\parallel$
and $\perp$ to represent projections onto these components.  In
particular, the components $(e_\parallel,e_\perp)$ of the ellipticity
vector $\vece$ have the values $e_\parallel=e\cos \ma$ and $e_\perp=-e\sin \ma$ in the unperturbed orbit.
The true anomaly $\nu,$ radius $r$, and azimuthal angle $\phi,$ and
the velocity components of the unperturbed orbit are, to first order in $e$,
\begin{eqnarray}
  \ma & = & \ma_0 + t = \tau_0,  \nonumber \\
  \nu & = & \ma - 2 e_\perp \nonumber \\
  \phi & = & \nu + \lop = \tau - 2e_\perp, \nonumber\\
  r & = & 1-e_\parallel, \nonumber \\
  v_r & = & -e_\perp \nonumber \\
  v_t & = & 1 + e_\parallel
            \label{eq:kepler}
\end{eqnarray}

The first element adopted for $\vecq$ is the semimajor axis $a$, with
shift derivable from the orbital energy $E=-1/2a.$
\begin{eqnarray}
  \Delta E & = &  \vecv \cdot \vecI + O(I^2) \\
  \quad \Rightarrow \quad \Delta a & = & 2\left(v_r I_r + v_\phi I_\phi\right) = -2e_\perp I_r + 2(1+e_\parallel) I_\phi.
  \label{eq:da}
\end{eqnarray}
The initial angular momentum $\mathbf{L} = \sqrt{1-e^2}\,\zhat \approx \zhat$ is altered by
\begin{eqnarray}
  \Delta \mathbf{L} & = & \vecr \times \vecI = (1-e_\parallel) I_\phi \zhat - (1-e_\parallel) I_z \phat \\
  \quad \Rightarrow \quad \Delta L_\parallel & = & -2e_\perp I_z
  \label{eq:dLpar}\\
  \Delta L_\perp & = & -(1-e_\parallel)  I_z
                  \label{eq:dLperp}
\end{eqnarray}
We adopt the $x$ and $y$ components of the orbital angular momentum
$(L_x, L_y)$ as two elements of $\vecq$ specifying the inclination and
ascending node $\Omega$ of the perturbed orbit.  These are a rotation
of $(\Delta L_\parallel, \Delta L_\perp)$ by the angle $\tau.$  The conversion from the first line above to the subsequent two makes use of the decomposition $\phat=\uhat_\perp + 2e_\perp \uhat_\parallel$ derivable from the value of $\phi$ in Equations~(\ref{eq:kepler}).

The eccentricity vector $\vece = \vecv \times (\vecr \times \vecv) - \rhat$ is altered by the impulse according to
\begin{eqnarray}
  \Delta\vece & = & (2\rhat - rv_r \phat) I_\phi - \phat I_r \\
\label{eq:depar}
  \quad \Rightarrow \quad \Delta e_\parallel & = &  -2e_\perp I_r + 2 I_\phi \\
\Delta e_\perp & = & -I_r - 3e_\perp I_\phi.
\label{eq:deperp}
\end{eqnarray}
We adopt $e_x$ and $e_y$ as two additional components of $\vecq,$ again related through a rotation by $\tau$ to the quantities in Equations~(\ref{eq:depar}) and (\ref{eq:deperp}).

The final element of $\vecq$ will be $\tau=\ma+\lop.$  The shift
induced by an impulse can be obtained by enforcing the
condition that the radius $r,$ or the azimuthal angle $\phi,$ must
remain constant during the impulse.  To find the latter, we need a
formula for $\nu(\ma)$ as a power series in $e.$  Standard formulae in
the literature \citep[\eg][]{tremaine} lead to:
\begin{eqnarray}
  \nu & = & \ma + 2e \sin \ma + \frac{5e^2}{4} \sin 2\ma + O(e^3) \\
  \label{eq:nue2}
    & = & \ma + 2(e\sin \ma) + \frac{5}{2} (e \sin \ma) (e \cos \ma) + O(e^3) 
\end{eqnarray}
We wish to find the first-order shift in $\phi=\nu + \lop$ and set it
to zero.  But the derivatives of $\ma$ and $\lop$ with impulse can become infinite.  Instead we introduce a perturbation $\Delta\tau = \Delta(\ma+\lop).$ We can now express several post-impact quantities as deviations from the unperturbed quantities:
\begin{eqnarray}
  \ma & = & \tau-\lop = \tau_0 -\lop + \Delta\tau \\
  e \sin \ma & = & e \sin (\tau_0 -\lop) + \Delta\tau \left[ e \cos  (\tau_0 -\lop)\right] \nonumber \\
\label{eq:esinM}
    & = & -e_{0\perp}-\Delta e_\perp + e_{0\parallel} \Delta\tau +O(e^2) \\
  e \cos \ma & = & e \cos (\tau_0 -\lop) - \Delta\tau \left[ e \sin  (\tau_0 -\lop)\right] \nonumber\\
\label{eq:ecosM}
    & = & e_{0\parallel}+\Delta e_\parallel + e_{0\perp} \Delta\tau +O(e^2) \\
\Rightarrow \quad \phi =\nu+\lop & = & \tau_0 + \Delta \tau + 2( -e_{0\perp}-\Delta e_\perp + e_{0\parallel} \Delta\tau )
                              + \frac{5}{2}\left( -e_{0\perp}\Delta e_\parallel-e_{0\parallel}\Delta e_\perp\right) + O(e^2).
\label{eq:phi}
\end{eqnarray}
Forcing $\phi$ to be unchanged during the impulse requires
\begin{eqnarray}
  \Delta\tau & = & \frac{5 e_\perp}{2} \Delta e_\parallel + \left(2-\frac{3e_\parallel}{2}\right) \Delta e_\perp \\
             & = & \left(-2+\frac{3e_\parallel}{2}\right) I_r - e_\perp I_\phi.
                   \label{eq:dtau0}
\end{eqnarray}
We have dropped the 0 subscripts on $e_\parallel,e_\perp$ at this point for brevity---they will refer to the values for the unperturbed orbit at the time of impulse, unless noted otherwise.

After the impulse, the mean anomaly $\ma$ advances at a rate of $a^{-3/2}t,$ meaning that an additional term $\Delta\tau = -3\Delta a (T-t_i)/2$ accrues by time $T$. The total perturbation of $\tau$ from the nominal value $T+\ma_0$ becomes
\begin{equation}
  \Delta\tau= \left[-2+\frac{3e_\parallel}{2} +3 e_\perp (T-t_i)\right] I_r  - \left[3(1+e_\parallel)(T-t_i) + e_\perp\right] I_\phi.
  \label{eq:dtau}
\end{equation}

We now have all the information we need to create the $\matA$ matrix.
Rotating to $\vecq=\{a, \tau, e_x, e_y, L_x, L_y\}$ and with $\vecI=\{I_r, I_\phi, I_z\}$ we have, to $O(e)$,
\begin{equation}
         \matA(e,T,t_i) = \left( \begin{array}{ccc}
 2e\sin\tau & 2 + 2e\cos\tau & 0 \\
 -2+e\left[1.5\cos\tau-3(T-t_i)\sin\tau\right] & e\sin\tau -3(1+e\cos\tau)(T-t_i) & 0\\
 \sin\tau+e\sin 2\tau & 2\cos\tau - 3e\sin^2\tau & 0\\
  -\cos\tau - 2e\sin^2\tau & 2\sin\tau-1.5e\sin 2\tau & 0 \\
0 & 0 & \sin\tau + 0.5 e \sin 2\tau \\
 0 & 0 &  -\cos\tau +e(3-\cos 2\tau)/2 \\
 \end{array}\right)
\end{equation}
We can now do the time integral and sum over impulse directions in \eqq{eq:Cqjk}, and average over $\tau$ to average over the orbital phase of the impulses. The diagonal elements of $\covm^q$ are
\begin{eqnarray}
  \Var(a) & = & 4T \nIphi \nonumber\\
  \Var(\tau) & = & 4T\nIr + 3T^3\nIphi \nonumber\\
  \Var(e_x) =  \Var(e_y) & = & \nIr/2+2\nIphi \nonumber\\
  \Var( L_x) =  \Var( L_y) & = & \nIz/2
\label{eq:Cq}
\end{eqnarray}
All of the off-diagonal terms (covariances) are either zero or $O\left(e\left\langle nI^2\right\rangle\right),$ and they result in observable consequences that are smaller by $\left\langle e^2 \right\rangle$ than the diagonal terms' contributions.

 \subsection{Astrometric and ranging shifts from elements shifts}
\label{sec:observe}
The radial coordinate of a Keplerian orbit is $r=a(1-e\cos E),$ and an expansion of the eccentric anomaly $E$ in powers of $e$ is
\begin{equation}
  \label{eq:Ee2}
  e \cos E  =  e\cos \ma - (e \sin \ma)^2 + O(e^3).
\end{equation}
From this we can derive the perturbations to $r$ as a function of our chosen basis for perturbations to the orbital elements.  Making use of Equations~(\ref{eq:ecosM}) and (\ref{eq:esinM}) we obtain
\begin{equation}
  \Delta r = (1-e\cos\tau) \Delta a + e\sin\tau \Delta\tau +  (-\cos\tau -2e\sin^2\tau) \Delta e_x + (-\sin\tau+e\sin 2\tau) \Delta e_y.
\end{equation}
If we square this expression, average over the orbital phase $\tau$ of the observation, and retain leading order in $e$, we get the variance of the radial coordinate from the unperturbed orbit.  All of the terms involving covariances between orbital elements either average to zero over an orbit, or are suppressed by $e^2$ relative to other terms.  We are left with
\begin{eqnarray}
  \sigma^2_r = \Var(r) & \approx & \Var(a) + \frac{1}{2}\Var(e_x) + \frac{1}{2}\Var(e_y) + \frac{\langle e^2\rangle}{2} \Var(\tau) \\
          & \approx & \frac{T}{2} \nIr + \left( 6T + \frac{3\langle e^2 \rangle}{2} T^3\right) \nIphi.
                      \label{eq:varr}
\end{eqnarray}
The second line substitutes the results from Equations~(\ref{eq:Cq}).  We retain the final term because, as $T>>1,$ this precession term that scales with $\langle e^2 \rangle T^3$ becomes comparable to or larger than the other terms that scale as $T$.  
The heliocentric azimuthal angle $\phi$ is given by \eqq{eq:phi}.  We can similarly square the perturbation and average over orbital phase and population factors to get
\begin{eqnarray}
\sigma^2_\phi =   \Var(\phi) & \approx & \Var(\tau) + 2\Var(e_x) + 2\Var(e_y) \\
             & \approx & 6T\nIr + \left(8T+3T^3\right) \nIphi.
                         \label{eq:varphi}
\end{eqnarray}
The heliocentric latitude (or equivalently, polar angle) has a deviation
\begin{eqnarray}
  \Delta\theta & = & -\Delta L_x \frac{x}{r} - \Delta L_y \frac{y}{r} \\
  \Rightarrow \quad \sigma^2_\theta = \Var(\theta) & \approx & \langle\cos^2\tau\rangle \Var(L_x) + \langle\sin^2\tau\rangle \Var(L_y) \\
               & \approx & \frac{T}{2} \nIz.
                           \label{eq:vartheta}
\end{eqnarray}

Equations~(\ref{eq:varr}), (\ref{eq:varphi}), and (\ref{eq:vartheta}) comprise our desired result.  Several characteristics are noteworthy:
\begin{itemize}
\item $r$ is in units of $a_0$ and $\theta, \phi$ are angles.  To convert their variances into linear displacements in the $r,\phi,z$ directions, multiply them all by $a^2,$ or equivalently scale the standard deviations by $a.$
\item The terms proportional to $T$ show typical diffusive growth
  $\sigma \propto \sqrt{T},$ and arise from random walks in the 5 constants of Keplerian motion.  They represent epicyclic alterations to the original orbit that grow in amplitude as $\sqrt{T}.$  The $T^3$ terms dominate $\Var(\phi)$ by the end of a full orbit, resulting from accumulating delays/advances $\Delta\tau$ in orbital phase $\ma$ as $a$ and the mean motion diffuse.  Thus it is the energy-changing impulses, namely $I_\phi,$ that cause the most Brownian motion.
\item The radial variance also becomes dominated by the $T^3$ term
  within 2~years at typical MBA values of $e$ (although
  nearly-circular orbits have $\Var(r)\propto T$)---this term arises from advances/delays in the radial epicycle phase.  This is the only part of the Brownian motion that whose leading order in $e$ is non-zero, \ie\ which we would have missed by assuming circular initial orbits.
\item The RMS vertical displacement grows strictly as $T^{1/2}.$
\item It will be generally true that $\Var(\phi) > \Var(r) > \Var(\theta),$ a trend exacerbated by the empirical observation (Figure~\ref{fig:rtz}) that impulses favor the $\phat$ direction.
\item There are no significant covariances among the $r,\theta,$ and $\phi$ displacements at the population-averaged level.
  \item For Earth-based observers, the deviations in $(r,\theta,\phi)$ from the nominal orbit will be some calculable, time-dependent, nearly-orthonormal matrix transformation of the heliocentric deviations derived above, so the results will not be grossly different aside from bringing the three directional components closer together in amplitude.
\end{itemize}

\section{Results and conclusions}
\label{sec:results}
We can now propagate the numerical estimates for $\left\langle nI_\gamma^2\right\rangle$ from Section~\ref{sec:impulse} through the analytical observational consequences in Section~\ref{sec:observe} to yield estimates of the unmodelled RMS shifts of MBAs from each region.  We assume the following parameters:
\begin{itemize}
\item The deviations are accumulated over a time $T=10$~yr.
\item The population has $\langle e^2 \rangle=0.03,$ which is the value for the cataloged MBAs of the Inner and Middle belts, and a slight overestimate for the Outer belt.
\item The $5\sigma$ impulse size $I_{\rm det}$, at which we deem it possible to isolate the effect of a single deflector on a tracer and hence usefully include that deflector's mass in the ephemeris model, is taken to be $I_{\rm det} = 10^{-8}v_c.$ In practice this will depend on the geometry of the impulse and on the quality of the tracer's observations.  But we find that changing $I_{\rm det}$ by a factor of 10 alters the noise from unmodeled impulses by $<20\%.$
  \item We will vary $\sigma_M,$ the RMS uncertainty on the mass of modelled asteroids, such that the number $N$ of MBAs  with $M>\sigma_M$ varies from 100 to 10,000.  The number of MBAs included in the ephemeris model would be $\approx N.$ 
\end{itemize}

\begin{figure}
  \centering
  \includegraphics[width=0.6\textwidth]{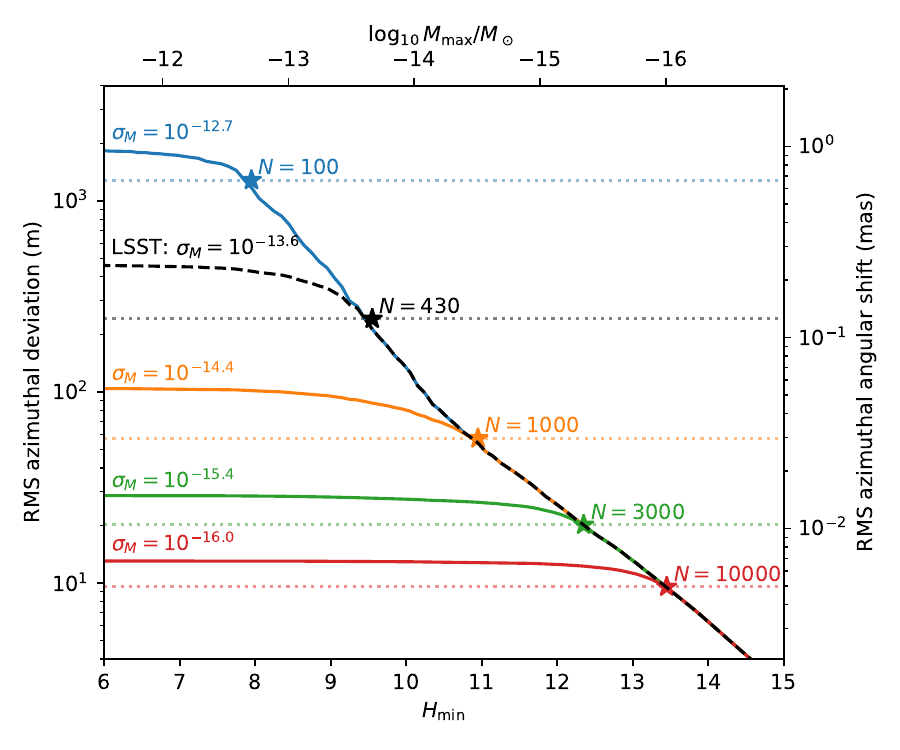}
  \caption{Each line shows the forecast of the standard deviation of the
    azimuthal position of a Middle Main Belt asteroid caused by discrete
    encounters over 10~years with absolute magnitude $H$ greater than the value on
    the (lower) $x$ axis, \ie\ the cumulative noise from all MBAs
    fainter (smaller) than the plotted $H_{\rm min.}$  The upper axis gives the $M_{\rm max}$ equivalent to $H_{\rm min}.$
.  The left axis gives RMS values in meters, the right
    the apparent shift in milliarseconds for a heliocentric observer.
    Each line is for a different assumption about the RMS uncertainty $\sigma_M$ in the masses of individual asteroids gained from high-$S/N$ observations of mutual encounters.
Each star marks the dividing point between the $N$ MBAs with $M>\sigma_M,$ and the smaller bodies with no useful measure of their mass ($M<\sigma_M$). The light dotted lines mark the level of azimuthal noise expected from the bodies beyond the most massive $N,$ \ie\ the RMS that an ephemeris model including $N$ bodies would not properly model. 
    The black dashed curve is a conservative estimate of what can be achieved using LSST data to constrain larger MBA's masses. Radial and vertical RMS positions are factors of $\approx 7$ and 45 lower than the azimuthal noise, respectively.}
  \label{fig:result}
\end{figure}

Figure~\ref{fig:result} plots the resulting noise level $\sigma_\phi=\sqrt{\Var(\phi)}$ of the calculation for the Middle regions of MBAs, showing both the linear RMS displacements on the left axis and the angular displacements on the right axis.  Each point on the curve is the $\sigma_\phi$ attributable to deflector MBAs that are smaller (fainter) than the $H$ value on the lower axis, or the mass on the top axis.  Thus the left-most point on each curve marks the total noise integrated over the full population of deflectors. Different color lines show the results for the values of $N$ and $\sigma_M$ as marked astride each line.  Notable features of these result are:
\begin{itemize}
\item \emph{The linear displacements are nearly identical for Inner,
    Middle, and Outer regions of the asteroid belt.}  The angular
  displacements vary slightly, inversely to their distance from the
  Sun.  Figure~\ref{fig:result} plots only the Middle MBAs'
  deviations, although the count $N$ of modelled MBAs is the sum for
  all three regions.
\item We plot the azimuthal noise.  \emph{The radial noise results
    have nearly identical scaling with $H$ and $\sigma_M$, but
    $\approx 7\times$ lower RMS. Likewise the vertical Brownian motion
    has the same shape but $\approx45\times$ lower than the azimuthal RMS.}
\item For longer time periods $T$, the azimuthal and radial noise will scale as
$T^{3/2},$ the vertical
noise as $T^{1/2}.$
\item The largest contributions to the unmodelled Brownian motion are
made by deflectors with mass and $H$ at the ``shoulder'' of each
curve, which will also correspond to the MBAs with masses $M\sim
\sigma_M$.  Above this mass, the noise per deflector is constant but
the deflectors become scarce at larger $M.$ Below this mass, $dN/dH$
is rising to fainter/smaller bodies, but the factor $M^2(H) dN/dH$
that determines the Brownian motion is dropping.
\end{itemize}

We take as representative of the present state of the art the case
where the $N=100$ most massive MBAs (aside from Ceres and Vesta) have
masses determined to $\sigma_M\approx 10^{-12.7} M_\odot$ by mutual
encounters.  Our calculations suggest then that a 10-year ephemeris
including these 100 MBAs will accrue RMS errors of $\approx 2$~km in
the azimuthal position, or 1~mas in ecliptic longitude.
RMS range and vertical ephemeris errors would be closer to 300 and 40~m, respectively.  We find that, roughly speaking, $\sigma_\phi$
drops with the number of modeled MBAs as $1/N.$

 In the next decade, LSST will significantly
increase the number $N$ of deflector MBAs with
usefully measured masses from mutual encounters by tracking hugely
more potential tracers and lowering $\sigma_M.$  The conservative estimate given in Section~\ref{sec:mbaprops} is that LSST will lower $\sigma_M$ to $\approx10^{-13.7},$ which our calculations indicate lowers the $\sigma_\phi$ values to $\approx500$~m or 0.25~mas.

We need to check whether our supposition that impulses $I>I_{\rm det}$
would stand out above the measurement noise of a single tracer remains
true once we include the noise induced by the other unmodelled
impulses.  Focusing on the effects of a tangential impulse of
$I_\phi=I_{\rm det}/\sqrt{2},$ \eqq{eq:dtau} predicts an orbital phase
shift of $\Delta\tau \approx 3TI_\phi$ at time $T$ past the impulse,
translating directly into a position shift $\Delta\phi$ of the same
size as per \eqq{eq:phi}.  Consider a survey of duration $T,$ and a
single impulse $I_{\rm det}$ applied at time $T/2.$ If
$I_\phi=10^{-8}/\sqrt{2}$ and $T=10$~yr, then
$\Delta\phi\approx35$~mas.  This is much larger than the
stochastic $\sigma_\phi\approx1$~mas expected from unmodelled masses at the
current state of knowledge.  For primarily radial or vertical
impulses, the astrometric signal will be smaller, but the Brownian
motion noise is lower too. 
We conclude that our choice of $I_{\rm
det}$ is not ruined by the presence of Brownian motion noise on the
tracer asteroids---impulses of this size will remain strongly detected.

\section{Consequences}
\label{sec:discuss}
\subsection{Implications for future inferences}
  
An important question is: will the positional noise due to unmodelled
MBA encounters degrade inferences being made with measurements of
tracer MBAs?  If the Brownian noise is well below other sources of measurement noise, then the answer must be ``no.''
The per-epoch uncertainties on solar-system object astrometry from Gaia\footnote{We take one Gaia transit as an ``epoch'' here.} are reported to be as low as 0.3~mas for very bright sources ($G=13$), rising to $\approx10$~mas at its $G=21$ limit where most detections lie \citep[][Figure 6]{gaiass3}.  The overall RMS residual to the orbit fits are 5~mas (this is in the ``along-track'' direction).  Thus the Brownian motion of $\approx1$~mas we estimate for $\phi$ (over a 10-year period) with current asteroid-mass knowledge would be important for the single-epoch error budget of the brighter Gaia MBAs.  With our forecasted LSST-era knowledge of asteroid masses ($\sigma_M\approx10^{-13.6},$ RMS azimuthal errors $\approx0.25$~mas), the situation will be improved in that only the brightest few MBAs will have per-epoch Brownian motion noise comparable to the per-epoch measurement noise.

For LSST, the measurement errors on MBA positions will have a floor of 1--2~mas per epoch set by refraction from atmospheric turbulence \citep{willow,trojans}.  Objects with apparent magnitude $m\gtrsim20$ will have larger per-epoch errors due to photon noise in the images.  Assuming that LSST inferences will be able to exploit LSST's own determinations of MBA masses, we see that on a per-epoch basis, the Brownian noise will always be subdominant to measurement errors.

An ambitious proposal by \citet{occultations} is to track $\approx10^6$ MBAs down to diameters of $\approx1$~km by watching them occult Gaia stars.  The timing of each occultation can locate the center of mass of the target MBA along its direction of motion, with most of the information available from objects yielding uncertainties of 50--100~m on this position.  The Brownian noise (with LSST-level asteroid knowledge) over a 10-year baseline will in this case be comparable to the per-epoch measurement errors for most observations.  Reducing the typical mass uncertainty of individual asteroids to $\sigma_M\approx 10^{-15.4} M_\odot$ would be needed ($N=3000$) to reduce the Brownian noise to negligible levels for this survey.

\emph{These per-epoch comparisons are, however, misleading.} The
Brownian noise is highly correlated between different epochs of
observation of a given asteroid, while the measurement errors are
generally uncorrelated, and therefore the impact on inferences is not
described solely by comparison of RMS values of each.  The covariance
structure of Brownian noise can work either to the benefit or the
detriment of the inference.  In an unfavorable case, the eigenvectors
of the Brownian covariance matrix with the largest uncertainties look similar to the signal of interest in the inference, \ie\ the derivative of the MBA's position with respect to a parameter.  If there is a single dominant eigenvector of the Brownian noise which looks just like our signal, then the per-epoch Brownian noise does not average down at all as we obtain $M$ observations, but the measurement noise drops as $\sqrt{M}.$  In a favorable case, all of the strong eigenvectors of the Brownian noise covariance are orthogonal to the signals of interest, and the Brownian noise causes no degradation of the inference, \ie\ its spatiotemporal structure cannot mimic that of the signal we seek.

Consider first the worst-case scenario, where the Brownian noise does
not average down at all across observations, but measurement errors
drop by a factor $\sqrt{M}.$   \citet{gaiafpr} report typically $\sim60$ epochs per asteroid in 66 months, suggesting $O(100)$ epochs per asteroid in the full 11-year observing period.  At Gaia's faint end, this means that Brownian noise with present asteroid knowledge would have comparable impact to the measurement errors, and dominate inferences at the bright end.  LSST will obtain several hundred epochs per asteroid.  With an effective factor 10--20 reduction in the total measurement noise, and assuming $\sigma_\phi\approx0.25$~mas Brownian noise, the latter would become important for the brighter tracer MBAs in LSST data, but inferences with the millions of fainter objects discovered by LSST would still be limited by photon noise.  The occultation program proposed by \citet{occultations} would obtain $<10$ measurements per decade for most MBAs and hence see less degradation of Brownian relative to measurement noise.

At the other extreme are cases where the dominant modes of Brownian noise are orthogonal to the effects of parameters of interest.  
If, for example, one is searching for the tidal force of a Planet X, this will be manifested primarily as a collective apsidal precession and a quadrupolar deviation of the orbit from its ellipse that does not evolve with time.  These are quite distinct from the random walk in $\Delta\tau$ that is the dominant effect of Brownian motion, and therefore Brownian noise will cause less degradation to the Planet X inference than the per-epoch $\sigma_\phi$ would suggest.  Another example would be a gravitational perturbation whose signature is primarily a nodal precession, which is a signal manifesting primarily as out-of-plane motions---we have seen that Brownian motion out of the orbital plane is $>40\times$ smaller than the $\phi$ component and grows at $T^{-1/2}$ instead of $T^{3/2},$ so the per-epoch $\sigma_\phi$ of Brownian noise will grossly overestimate its effect on the inference.

The algorithms used to make inferences based on range or astrometric
data will need to be changed when Brownian motion becomes a
significant source of error, to accomodate the fact that positional
errors on a given tracer are highly correlated between epochs.  The
full covariance matrix of the tracer's observations will be needed
rather than typical practice of considering measurements as
independent.  A small modification to our $\matB$ matrices would be
needed to calculate the covariance of the Brownian noise between
observations at different times $T$ and $T^\prime.$  The Brownian
motion should be fairly uncorrelated between different MBAs, because
the varying encounter geometries randomize the influence of any
particular deflector's mass error on the trajectories of the tracers
it influences.

A further difficulty in propagating unmodelled deflections into
inferences is that the probability distribution of the position shifts
will be non-Gaussian, because the RMS is dominated by the
contributions of relatively few encounters with the largest unmodelled
asteroids.  The central limit theorem will not be applicable over
decade timescales.

A better approach, however, would probably be to increase the number
$N$ of MBAs considered as active bodies in the ephemeris fitting,
leaving their masses as free parameters whether or not we expect them
to be constrained to $S/N\gtrsim1$ by the data.  Although the extreme
case of having $N=O(10^6)$ active bodies' masses as free parameters in
the ephemeris fit would be a substantial computational challenge,
having $N=10^3$--$10^4$ should be manageable.  The MBA-to-MBA perturbations can be adequately calculated using Newtonian dynamics instead of a fully relativistic treatment.  In this approach, all of the covariances between different observations of the same tracer, and between tracers, will be captured in the analysis for deflections due to these $N$ deflectors.  If we \emph{choose $N$ to be large enough that the Brownian noise from the unmodelled MBAs is negligible to the error budget,} then we need not explicitly propagate the Brownian noise.

Figure~\ref{fig:result} shows us what $N$ would be required: the noise due to unmodelled deflectors is given by the point where each curve for a given $N$ departs from the upper-right envelope of the curves.   Increasing the number of MBAs in the ephemeris from $N=100\rightarrow10^3\rightarrow10^4$ decreases the RMS displacement in 10~yrs due to the unmodelled MBAs from $\sigma_\phi=1300\rightarrow60\rightarrow10$~m, with the radial and vertical RMS being substantially lower.  These correspond to 0.7, 0.03, and 0.005~mas, respectively.  Incorporating $N=1000$ MBAs into the ephemeris model would render noise from the rest of the asteroids unimportant for nearly all uses of Gaia or LSST MBA data for inference.  For occultation-based positions, one would probably want to have the $\approx3000$ most massive MBAs as active bodies with masses as free parameters in the ephemeris model in order to render encounters from unmodelled bodies as unimportant.

\subsection{Comparison to radiation forces}
As we consider smaller MBAs as test particles, radiation pressure becomes larger relative to gravitational forces.  It is interesting to compare the effects of radiation pressure to those of Brownian motion to see when the former are a bigger source of noise than the latter, or contrarily to see whether Brownian motion interferes with attempts to measure radiation forces on MBAs.  At first order, radiation pressure from the Sun and the reaction from reflected sunlight cause a constant radial force on the MBA.  The ratio of this force to the gravitational force for an asteroid of radius $R,$ geometric albedo $q$, and density $\rho$ is $(1+q)\Psi,$ where
\begin{equation}
 \Psi \equiv \frac{F_{\rm rad}}{F_{\rm grav}} = \frac{\pi R^2 L_\odot/4\pi c}{4\pi GM_\odot R^3\rho/3}
   = \frac{3}{16\pi} \frac{L_\odot}{GM_\odot c R \rho} \approx
   3\times10^{-10} \times 10^{0.2(H-16)},
\end{equation}
where the last expression adopts our standard values $q=0.25,
\rho=2500\,{\rm kg}\, {\rm m}^{-3}.$ The primary observational consequence of
this constant radial force is a time-independent, $H$-dependent 
shift $\Delta r = -(1+q)\Psi a$ of the radial position a body should
have at a given period.  
This does not resemble
any expected effect of Brownian motion, and also very difficult to
measure since radar ranging to MBAs is infeasible.  More important for
observations is the
Yarkovsky effect, in which anisotropic re-emission of the absorbed
radiation leads to a net force that is a fraction $g_Y(1-q)$ of the incident radiation pressure.  The Yarkovsky effect has been detected for several NEAs but no individual MBAs, with values $g_Y \approx 0.1$ \citep[Yarkovsky effect and data are review by][]{yarkovsky}.  This is also consistent with detections of the Yarkovsky effect by examining the $\Delta a$ dependence on $H$ of members of asteroid \emph{families} in the Main Belt \citep[\eg][]{karin}.

 The Yarkovsky effect has in common with Brownian noise that it has the strongest observable effects via changes to $a$ and the mean-motion rates, and that these changes can be of varying sign and amplitude for a given asteroid.  In the case of the Yarkovsky effect, this variability arises because of the random orientations of spin axes, and differences in spin rates, shapes, and surface properties.  
 
Taking a case where the Yarkovsky force is in the prograde azimuthal direction, and using our system of units with $a=1$ and $v_c=1,$ this leads to a continuous power $dE/dt\approx g_Y\Psi$ which in turn implies $da/dt\approx2g_Y\Psi.$   The mean motion rate is hence $1-3 g_Y\Psi t$ and the accumulated shift in mean anomaly over time $T$ amounts to $\Delta\phi = -3 g_Y \Psi  T^2/2.$   In this case, a rough comparison of the accumulated $\Delta\phi$ from the Yarkovsky effect to the $\sigma_\phi$ of noise from unmodelled Brownian motion, after a time interval $T$, is
\begin{equation}
\frac{\textrm{Yarkovsky}}{\textrm{Brownian}} = 
   \frac{\Delta\phi}{\sigma_\phi} \approx \frac{ 3 g_Y \Psi
     T^2/2}{\sqrt{3 T^3 \nIphi}} \approx \sqrt{T/10\,{\rm yr}} \times 10^{0.2(H-20.2)}
     \label{eq:yarkovsky}
\end{equation}
if we assume $\nIphi$ from the current knowledge of asteroid masses.   There is thus a break point near $H=20$, or $\approx100$~m diameter: for MBAs larger than this, the ephemeris uncertainties from the Yarkovsky effect will be smaller than those from mutual gravitational encounters on decade timescales. This includes all MBAs detected by Gaia and nearly all detectable by LSST.  Below this size, the Yarkovsky displacement grows larger in a decade and has a better chance of being detectable in the presence of Brownian motion.  Distinguishing the steady quadratic growth of the Yarkovsky $\Delta\phi$ from the random-walk growth of Brownian motion to the same final size requires observational S/N and time resolution significantly better than is needed for detecting either one individually.

When Yarkovsky forces on MBAs are estimated using the $H$-dependent
dispersion of $a$ for asteroid families, the effective ``observation''
time period is typically several Myr.  Over these long time scales,
\eqq{eq:yarkovsky} shows that the temporally coherent Yarkovsky effect
can dominate over Brownian motion for bodies up to 10's or 100's of km
in diameter, which is why the technique is successful.

\subsection{Summary}

In conclusion, the Brownian motion from unmodelled or mis-modelled
mutual impulses between MBAs causes RMS azimuthal displacements of
$\approx2$~km (or 1 mas in angular position) over 10 years, with
current knowledge of the masses of large asteroids.  This is large
enough to become a significant part of the error budget for many
inferences from Gaia MBA astrometry.  The knowledge on large-MBA
masses likely to be gained from LSST tracking of MBAs should reduce
this by a factor of $\ge4,$ at which point the Brownian noise will be
comparable to Gaia or LSST measurement uncertainties only for the
brighter targets of each survey.  In all cases, the Brownian motion is
$\approx7\times$ smaller in the radial direction and $\approx45\times$
in the vertical direction.  The RMS azimuthal and radial displacements
grow with time as $T^{3/2}$, and the vertical RMS as $T^{1/2}.$

Making inferences from MBA astrometry becomes more complicated when
Brownian motion is a significant contributor to the error budget, but
a good path forward is to do all model fitting by including the
$\approx1000$ largest MBAs as active bodies in the ephemeris model and
leaving their masses as free parameters.  This will constrain the
masses of these bodies to the full extent that the data allow, and
will properly propagate any remaining uncertainty in those masses into
the other inferences that one is trying to make.  The noise
contributed by the remaining millions of unmodelled MBAs would be just
$\approx60$~m RMS in the azimuthal direction (about 30~$\mu$as), and a
few $\mu$as or less in the other axes, small enough to ignore entirely
for the accuracy of imaging astrometry for the foreseeable future.
Occultation-based astrometry is much more accurate (but has fewer possible measurements per tracer), and would be significantly impacted by Brownian motion until $\gtrsim1000$ MBAs have mass estimates at modest $S/N$ level, although the precise impact will depend on whether the signals one is seeking are parallel or orthogonal to the kind of variations caused by Brownian motion.  For objects $\gtrsim100$~m in diameter, uncertainties from mutual encounters are larger than uncertainties from the Yarkovsky effect over decade timescales, at present levels of knowledge of asteroid masses.

\begin{acknowledgments}
  This work was supported by NSF grant AST-2205808.  GMB is grateful for
  the thoughts and efforts of Negin Najafi, Daniel Gomes, Matt Holman,
  David Trilling, and Will Grundy.
  This research has made use of data and/or services provided by the International Astronomical Union's Minor Planet Center. 
\end{acknowledgments}

\newpage
\bibliographystyle{aasjournal}
\bibliography{references}

\begin{thebibliography}{}
\expandafter\ifx\csname natexlab\endcsname\relax\def\natexlab#1{#1}\fi
\providecommand{\url}[1]{\href{#1}{#1}}
\providecommand{\dodoi}[1]{doi:~\href{http://doi.org/#1}{\nolinkurl{#1}}}
\providecommand{\doeprint}[1]{\href{http://ascl.net/#1}{\nolinkurl{http://ascl.net/#1}}}
\providecommand{\doarXiv}[1]{\href{https://arxiv.org/abs/#1}{\nolinkurl{https://arxiv.org/abs/#1}}}

\bibitem[{{Baer} \& {Chesley}(2017)}]{baer}
{Baer}, J., \& {Chesley}, S.~R. 2017, \aj, 154, 76,
  \dodoi{10.3847/1538-3881/aa7de8}

\bibitem[{{Bernstein} {et~al.}(2025){Bernstein}, {Najafi}, \& {Gomes}}]{negin}
{Bernstein}, G., {Najafi}, N., \& {Gomes}, D. 2025, \psj, in preparation

\bibitem[{{Fienga} {et~al.}(2020{\natexlab{a}}){Fienga}, {Avdellidou}, \&
  {Hanu{\v{s}}}}]{inpop}
{Fienga}, A., {Avdellidou}, C., \& {Hanu{\v{s}}}, J. 2020{\natexlab{a}},
  \mnras, 492, 589, \dodoi{10.1093/mnras/stz3407}

\bibitem[{{Fienga} {et~al.}(2020{\natexlab{b}}){Fienga}, {Di Ruscio}, {Bernus},
  {Deram}, {Durante}, {Laskar}, \& {Iess}}]{inpopP9}
{Fienga}, A., {Di Ruscio}, A., {Bernus}, L., {et~al.} 2020{\natexlab{b}}, \aap,
  640, A6, \dodoi{10.1051/0004-6361/202037919}

\bibitem[{{Fortino} {et~al.}(2021){Fortino}, {Bernstein}, {Bernardinelli},
  {Aguena}, {Allam}, {Annis}, {Bacon}, {Bechtol}, {Bhargava}, {Brooks},
  {Burke}, {Carretero}, {Choi}, {Costanzi}, {Costa}, {Pereira}, {De Vicente},
  {Desai}, {Doel}, {Drlica-Wagner}, {Eckert}, {Eifler}, {Evrard}, {Ferrero},
  {Frieman}, {Garc{\'\i}a-Bellido}, {Gazta{\~n}aga}, {Gerdes}, {Gruendl},
  {Gschwend}, {Gutierrez}, {Hartley}, {Hinton}, {Hollowood}, {Honscheid},
  {James}, {Jarvis}, {Kent}, {Kuehn}, {Kuropatkin}, {Maia}, {Marshall},
  {Menanteau}, {Miquel}, {Morgan}, {Myles}, {Ogando}, {Palmese},
  {Paz-Chinch{\'o}n}, {Plazas}, {Roodman}, {Rykoff}, {Sanchez}, {Santiago},
  {Scarpine}, {Schubnell}, {Serrano}, {Sevilla-Noarbe}, {Smith}, {Suchyta},
  {Tarle}, {To}, {Tucker}, {Varga}, {Walker}, {Weller}, {Wester}, \& {DES
  Collaboration}}]{willow}
{Fortino}, W.~F., {Bernstein}, G.~M., {Bernardinelli}, P.~H., {et~al.} 2021,
  \aj, 162, 106, \dodoi{10.3847/1538-3881/ac0722}

\bibitem[{{Gaia Collaboration} {et~al.}(2023){Gaia Collaboration}, {David},
  {Mignard}, {Hestroffer}, {Tanga}, {Spoto}, {Berthier}, {Pauwels}, {Roux},
  {Barbier}, {Cellino}, {Carry}, {Delbo}, {Dell'Oro}, {Fouron}, {Galluccio},
  {Klioner}, {Mary}, {Muinonen}, {Ordenovic}, {Oreshina-Slezak}, {Panem},
  {Petit}, {Portell}, {Brown}, {Thuillot}, {Vallenari}, {Prusti}, {de Bruijne},
  {Arenou}, {Babusiaux}, {Biermann}, {Creevey}, {Ducourant}, {Evans}, {Eyer},
  {Guerra}, {Hutton}, {Jordi}, {Lammers}, {Lindegren}, {Luri}, {Randich},
  {Sartoretti}, {Smiljanic}, {Walton}, {Bailer-Jones}, {Bastian}, {Cropper},
  {Drimmel}, {Katz}, {Soubiran}, {van Leeuwen}, {Audard}, {Bakker}, {Blomme},
  {Casta{\~n}eda}, {De Angeli}, {Fabricius}, {Fouesneau}, {Fr{\'e}mat},
  {Guerrier}, {Masana}, {Messineo}, {Nicolas}, {Nienartowicz}, {Pailler},
  {Panuzzo}, {Riclet}, {Seabroke}, {Sordo}, {Th{\'e}venin}, {Gracia-Abril},
  {Teyssier}, {Altmann}, {Benson}, {Burgess}, {Busonero}, {Busso},
  {C{\'a}novas}, {Cheek}, {Clementini}, {Damerdji}, {Davidson}, {de Teodoro},
  {Delchambre}, {Fraile Garcia}, {Garabato}, {Garc{\'\i}a-Lario}, {Garralda
  Torres}, {Gavras}, {Haigron}, {Hambly}, {Harrison}, {Hatzidimitriou},
  {Hern{\'a}ndez}, {Hodgkin}, {Holl}, {Jamal}, {Jordan}, {Krone-Martins},
  {Lanzafame}, {L{\"o}ffler}, {Lorca}, {Marchal}, {Marrese}, {Moitinho},
  {Nu{\~n}ez Campos}, {Osborne}, {Pancino}, {Recio-Blanco}, {Riello},
  {Rimoldini}, {Robin}, {Roegiers}, {Sarro}, {Schultheis}, {Siopis}, {Smith},
  {Sozzetti}, {Utrilla}, {van Leeuwen}, {Weingrill}, {Abbas},
  {{\'A}brah{\'a}m}, {Abreu Aramburu}, {Aerts}, {Altavilla}, {{\'A}lvarez},
  {Alves}, {Anderson}, {Antoja}, {Baines}, {Baker}, {Balog}, {Barache},
  {Barbato}, {Barros}, {Barstow}, {Bartolom{\'e}}, {Bashi}, {Bauchet},
  {Baudeau}, {Becciani}, {Bedin}, {Bellas-Velidis}, {Bellazzini}, {Beordo},
  {Berihuete}, {Bernet}, {Bertolotto}, {Bertone}, {Bianchi}, {Binnenfeld},
  {Blazere}, {Boch}, {Bombrun}, {Bouquillon}, {Bragaglia}, {Braine},
  {Bramante}, {Breedt}, {Bressan}, {Brouillet}, {Brugaletta}, {Bucciarelli},
  {Butkevich}, {Buzzi}, {Caffau}, {Cancelliere}, {Cannizzo}, {Carballo},
  {Carlucci}, {Carnerero}, {Carrasco}, {Carretero}, {Carton}, {Casamiquela},
  {Castellani}, {Castro-Ginard}, {Cesare}, {Charlot}, {Chemin}, {Chiaramida},
  {Chiavassa}, {Chornay}, {Collins}, {Contursi}, {Cooper}, {Cornez}, {Crosta},
  {Crowley}, {Dafonte}, {de Laverny}, {De Luise}, {De March}, {de Souza}, {de
  Torres}, {del Peloso}, \& {Delgado}}]{gaiafpr}
{Gaia Collaboration}, {David}, P., {Mignard}, F., {et~al.} 2023, \aap, 680,
  A37, \dodoi{10.1051/0004-6361/202347270}

\bibitem[{{Goffin}(2014)}]{goffin}
{Goffin}, E. 2014, \aap, 565, A56, \dodoi{10.1051/0004-6361/201322766}

\bibitem[{{Gomes} \& {Bernstein}(2025)}]{occultations}
{Gomes}, D. C.~H., \& {Bernstein}, G.~M. 2025, \psj, 6, 19,
  \dodoi{10.3847/PSJ/ad9f5b}

\bibitem[{{Gomes} {et~al.}(2023){Gomes}, {Murray}, {Gomes}, {Holman}, \&
  {Bernstein}}]{trojans}
{Gomes}, D. C.~H., {Murray}, Z., {Gomes}, R. C.~H., {Holman}, M.~J., \&
  {Bernstein}, G.~M. 2023, \psj, 4, 66, \dodoi{10.3847/PSJ/acc7a2}

\bibitem[{{Holman} \& {Payne}(2016)}]{holmanP9}
{Holman}, M.~J., \& {Payne}, M.~J. 2016, \aj, 152, 94,
  \dodoi{10.3847/0004-6256/152/4/94}

\bibitem[{{Konopliv} {et~al.}(2014){Konopliv}, {Asmar}, {Park}, {Bills},
  {Centinello}, {Chamberlin}, {Ermakov}, {Gaskell}, {Rambaux}, {Raymond},
  {Russell}, {Smith}, {Tricarico}, \& {Zuber}}]{Vesta}
{Konopliv}, A.~S., {Asmar}, S.~W., {Park}, R.~S., {et~al.} 2014, \icarus, 240,
  103, \dodoi{10.1016/j.icarus.2013.09.005}

\bibitem[{{Konopliv} {et~al.}(2018){Konopliv}, {Park}, {Vaughan}, {Bills},
  {Asmar}, {Ermakov}, {Rambaux}, {Raymond}, {Castillo-Rogez}, {Russell},
  {Smith}, \& {Zuber}}]{Ceres}
{Konopliv}, A.~S., {Park}, R.~S., {Vaughan}, A.~T., {et~al.} 2018, \icarus,
  299, 411, \dodoi{10.1016/j.icarus.2017.08.005}

\bibitem[{{LSST Science Collaboration} {et~al.}(2009){LSST Science
  Collaboration}, {Abell}, {Allison}, {Anderson}, {Andrew}, {Angel}, {Armus},
  {Arnett}, {Asztalos}, {Axelrod}, {Bailey}, {Ballantyne}, {Bankert},
  {Barkhouse}, {Barr}, {Barrientos}, {Barth}, {Bartlett}, {Becker}, {Becla},
  {Beers}, {Bernstein}, {Biswas}, {Blanton}, {Bloom}, {Bochanski}, {Boeshaar},
  {Borne}, {Bradac}, {Brandt}, {Bridge}, {Brown}, {Brunner}, {Bullock},
  {Burgasser}, {Burge}, {Burke}, {Cargile}, {Chandrasekharan}, {Chartas},
  {Chesley}, {Chu}, {Cinabro}, {Claire}, {Claver}, {Clowe}, {Connolly}, {Cook},
  {Cooke}, {Cooray}, {Covey}, {Culliton}, {de Jong}, {de Vries}, {Debattista},
  {Delgado}, {Dell'Antonio}, {Dhital}, {Di Stefano}, {Dickinson}, {Dilday},
  {Djorgovski}, {Dobler}, {Donalek}, {Dubois-Felsmann}, {Durech},
  {Eliasdottir}, {Eracleous}, {Eyer}, {Falco}, {Fan}, {Fassnacht}, {Ferguson},
  {Fernandez}, {Fields}, {Finkbeiner}, {Figueroa}, {Fox}, {Francke}, {Frank},
  {Frieman}, {Fromenteau}, {Furqan}, {Galaz}, {Gal-Yam}, {Garnavich},
  {Gawiser}, {Geary}, {Gee}, {Gibson}, {Gilmore}, {Grace}, {Green}, {Gressler},
  {Grillmair}, {Habib}, {Haggerty}, {Hamuy}, {Harris}, {Hawley}, {Heavens},
  {Hebb}, {Henry}, {Hileman}, {Hilton}, {Hoadley}, {Holberg}, {Holman},
  {Howell}, {Infante}, {Ivezic}, {Jacoby}, {Jain}, {R}, {Jedicke}, {Jee},
  {Garrett Jernigan}, {Jha}, {Johnston}, {Jones}, {Juric}, {Kaasalainen},
  {Styliani}, {Kafka}, {Kahn}, {Kaib}, {Kalirai}, {Kantor}, {Kasliwal},
  {Keeton}, {Kessler}, {Knezevic}, {Kowalski}, {Krabbendam}, {Krughoff},
  {Kulkarni}, {Kuhlman}, {Lacy}, {Lepine}, {Liang}, {Lien}, {Lira}, {Long},
  {Lorenz}, {Lotz}, {Lupton}, {Lutz}, {Macri}, {Mahabal}, {Mandelbaum},
  {Marshall}, {May}, {McGehee}, {Meadows}, {Meert}, {Milani}, {Miller},
  {Miller}, {Mills}, {Minniti}, {Monet}, {Mukadam}, {Nakar}, {Neill}, {Newman},
  {Nikolaev}, {Nordby}, {O'Connor}, {Oguri}, {Oliver}, {Olivier}, {Olsen},
  {Olsen}, {Olszewski}, {Oluseyi}, {Padilla}, {Parker}, {Pepper}, {Peterson},
  {Petry}, {Pinto}, {Pizagno}, {Popescu}, {Prsa}, {Radcka}, {Raddick},
  {Rasmussen}, {Rau}, {Rho}, {Rhoads}, {Richards}, {Ridgway}, {Robertson},
  {Roskar}, {Saha}, {Sarajedini}, {Scannapieco}, {Schalk}, {Schindler}, \&
  {Schmidt}}]{lsstbook}
{LSST Science Collaboration}, {Abell}, P.~A., {Allison}, J., {et~al.} 2009,
  arXiv e-prints, arXiv:0912.0201, \dodoi{10.48550/arXiv.0912.0201}

\bibitem[{{Mariani} {et~al.}(2023){Mariani}, {Fienga}, {Minazzoli},
  {Gastineau}, \& {Laskar}}]{inpopgraviton}
{Mariani}, V., {Fienga}, A., {Minazzoli}, O., {Gastineau}, M., \& {Laskar}, J.
  2023, \prd, 108, 024047, \dodoi{10.1103/PhysRevD.108.024047}

\bibitem[{Nesvorný \& Bottke(2004)}]{karin}
Nesvorný, D., \& Bottke, W.~F. 2004, Icarus, 170, 324,
  \dodoi{https://doi.org/10.1016/j.icarus.2004.04.012}

\bibitem[{{Park} {et~al.}(2021){Park}, {Folkner}, {Williams}, \&
  {Boggs}}]{de440}
{Park}, R.~S., {Folkner}, W.~M., {Williams}, J.~G., \& {Boggs}, D.~H. 2021,
  \aj, 161, 105, \dodoi{10.3847/1538-3881/abd414}

\bibitem[{{Pitjeva} \& {Pitjev}(2018)}]{pitjeva}
{Pitjeva}, E.~V., \& {Pitjev}, N.~P. 2018, Astronomy Letters, 44, 554,
  \dodoi{10.1134/S1063773718090050}

\bibitem[{{Tanga} {et~al.}(2023){Tanga}, {Pauwels}, {Mignard}, {Muinonen},
  {Cellino}, {David}, {Hestroffer}, {Spoto}, {Berthier}, {Guiraud}, {Roux},
  {Carry}, {Delbo}, {Dell'Oro}, {Fouron}, {Galluccio}, {Jonckheere}, {Klioner},
  {Lefustec}, {Liberato}, {Ord{\'e}novic}, {Oreshina-Slezak}, {Penttil{\"a}},
  {Pailler}, {Panem}, {Petit}, {Portell}, {Poujoulet}, {Thuillot}, {Van
  Hemelryck}, {Burlacu}, {Lasne}, \& {Managau}}]{gaiass3}
{Tanga}, P., {Pauwels}, T., {Mignard}, F., {et~al.} 2023, \aap, 674, A12,
  \dodoi{10.1051/0004-6361/202243796}

\bibitem[{{Thoss} \& {Burkert}(2024)}]{pbh}
{Thoss}, V., \& {Burkert}, A. 2024, arXiv e-prints, arXiv:2409.04518,
  \dodoi{10.48550/arXiv.2409.04518}

\bibitem[{{Tremaine}(2023)}]{tremaine}
{Tremaine}, S. 2023, {Dynamics of Planetary Systems} (Princeton University
  Press)

\bibitem[{{Vokrouhlick{\'y}} {et~al.}(2015){Vokrouhlick{\'y}}, {Bottke},
  {Chesley}, {Scheeres}, \& {Statler}}]{yarkovsky}
{Vokrouhlick{\'y}}, D., {Bottke}, W.~F., {Chesley}, S.~R., {Scheeres}, D.~J.,
  \& {Statler}, T.~S. 2015, in Asteroids IV, ed. P.~{Michel}, F.~E. {DeMeo}, \&
  W.~F. {Bottke} (University of Arizona Press), 509--531,
  \dodoi{10.2458/azu_uapress_9780816532131-ch027}

\end{thebibliography}

\end{document}